\documentclass[a4,aps,twocolumn,showpacs,preprintnumbers,qamslatex,amsmath,amssymb,longbibliography]{revtex4-1}
\usepackage[cp1251]{inputenc}
\usepackage[T2A]{fontenc}
\usepackage{fancyhdr}
\usepackage[colorlinks]{hyperref}
\usepackage{graphicx}
\usepackage{verbatim}
\usepackage{indentfirst}
\usepackage{latexsym}
\usepackage{dsfont}
\usepackage{epstopdf}

\begin{document}

\title{First-principle construction of U(1) symmetric matrix product states}

\author{Mykhailo V. Rakov}

\affiliation{Faculty of Physics, Kyiv National University, 64/13 Volodymyrska st., 01601 Kyiv, Ukraine}

\begin{abstract}
The algorithm to calculate the sets of symmetry sectors for virtual indices of
U(1) symmetric matrix product states (MPS) is described.
Principal differences between open (OBC) and periodic (PBC) boundary conditions are stressed,
and the extension of PBC MPS algorithm to projected entangled pair states (PEPS) is outlined.
\end{abstract}

\maketitle

\section{Introduction}
\label{intro}

In the last few decades extremely low temperatures have become routinely feasible, and various quantum systems have become possible to manipulate at such temperatures in mesoscopic and microscopic scale. The thermal degrees of freedom in these systems are `frozen', however, they may be quantum mechanically strongly correlated. 

One kind of such systems are low-dimensional quantum magnets, such as: quasi-one-dimensional compounds based on ions of various metals~\cite{Mikeska2004}; spin ladders~\cite{Brenig1995}; high-temperature superconductors and frustrated antiferromagnets~\cite{PhysRevB.82.155138}. Another kind are ultracold atomic gases: the simplest example is Bose-Einstein condensate (BEC)~\cite{RevModPhys.71.463}; however, quantum models with topological order~\cite{Kitaev2006,Wen1990} have also been realized with cold atoms~\cite{PhysRevLett.91.090402,Micheli2006}.

Remarkably, low-dimensional quantum magnets can be modeled by various spin Hamiltonians (e.g., a variety of Heisenberg Hamiltonians). Frustrated antiferromagnetic state can be partly understood on the basis of the 2D Hubbard model in the $t-J$ limit~\cite{auerbach} for spin-1/2 fermions. The dynamics of a Bose gas exposed to an optical lattice potential~\cite{PhysRevLett.81.3108,PhysRevLett.95.040402,Greiner2002} realizes a variety of Bose-Hubbard (BH) models at various conditions~\cite{essler}. But remarkably, the Hubbard models also reduce to various spin models in certain limits. As a consequence, quantum spin models are very important for understanding of the physics of low-dimensional systems at low temperatures.

The interactions between particles in the low-dimensional quantum systems can be versatile, and the `strength' of each interaction is determined by so-called control parameters that can be tuned experimentally. The system in its ground state is characterized by presence of specific correlation functions in different ranges of control parameters (these ranges are called `quantum phases', and the transitions between them are called quantum phase transitions (QPTs)). Therefore it is important to calculate the properties of the ground state of the system as functions of control parameters (sometimes also the properties of the leading excitations are useful).

It often happens that the Hamiltonian of a one-dimensional (1D) quantum system is U(1) symmetric (like Hubbard or Heisenberg Hamiltonians). In this case any eigenstate of this system is also U(1) symmetric due to Mermin-Wagner theorem~\cite{PhysRevLett.17.1133} even in the thermodynamic limit. This holds for any temperature, including absolute zero. The same also holds for two-dimensional (2D) systems for finite temperatures. Therefore it is natural to utilize such a property of an eigenstate if is to be determined in the variational algorithms.

The states of a quantum system are typically represented by an ansatz called projected entangled pair states (PEPS)~\cite{Orus2014}. The states of a one-dimensional system are thus represented as the simplest case of PEPS - the matrix product states (MPS)~\cite{Schollwoeck2011}.

Let us restrict (without loss of generality) our discussion to spin systems. I stress that I consider only {\it finite} systems. The U(1) symmetric eigenstate of a spin system has a well-defined total spin projection $S_z$. To ensure this, the elements of the MPS/PEPS tensors must have specific values. More precisely, the tensor elements whose positions do not satisfy certain `selection rules' equal zero. Therefore, it is desirable to represent the MPS/PEPS in an appropriate form from the outset, from both physical and numerical (computational) reasons.

The block structure of each \textit{local} PEPS (MPS in 1D) tensor is determined by Wigner-Eckart theorem~\cite{PhysRevB.83.115125}. However, the Wigner-Eckart theorem concerns only quantum numbers (spin projections for a spin system) but not their degeneracies. Furthermore, the set of quantum numbers at \textit{each} virtual (bond) dimension and their degeneracies are not dictated by the symmetry. The task is to calculate them, and it appears that in case of MPS for open boundary conditions (OBC) these sets can be calculated from first principles.

In this manuscript I outline the general way how the U(1) symmetric MPS for OBC may be constructed. The relation to well-known OBC implementations (e.g.,~\cite{McCulloch2007}) will be stressed. Analogous idea can be used for construction of the U(1) symmetric MPS for periodic boundary conditions (PBC), as well as for U(1) symmetric PEPS. This material goes beyond the scope of the current manuscript. Only principal differences between OBC and PBC in 1D will be shown here, as well as between 1D MPS and 2D PEPS (on a square lattice) with OBC.

\section{Origin of the symmetry sectors}
Let us denote the local physical index (a spin projection of a spin at site $i$) as $s_i$. The local MPS tensor $M_{a_{i-1},a_i}^{s_i}$ at position $i$ is rank-3 with $a_{i-1}$ and $a_i$ the left and the right virtual index, respectively. Analogously, the local PEPS tensor $M_{a_{i-1,(j)},a_{i,(j)},a_{(i),j-1},a_{(i),j}}^{s_{i,j}}$ at position $\{i,j\}$ is rank-5 with $a_{i-1,(j)},a_{i,(j)},a_{(i),j-1},a_{(i),j}$ the left, the right, the bottom and the top virtual indices.

Accordingly, the MPS for a 1D system of $N$ spins has the form

\begin{equation}
|\psi\rangle=\sum_{\{s\}} {\rm Tr}(M^{[1],s_1} M^{[2],s_2} \cdots M^{[N],s_N}) \, |s_1 \cdots s_N \rangle
\end{equation}
while the PEPS for a $N \times N$ spin square lattice is

$$
|\psi\rangle=\sum_{\{s\}} {\rm tTr}(M^{[1,1],s_{1,1}} M^{[1,2],s_{1,2}} \cdots
\cdot M^{[N,N-1],s_{N,N-1}} \cdot \nonumber
$$

\begin{equation}
\,\,\,\,\,\,\,\,\,\,\,\,\,\,\,\,\,\,\,\,\,\,\,\,\,\,\cdot M^{[N,N],s_{N,N}}) \, |s_{1,1} s_{1,2} \cdots s_{N,N-1} s_{N,N} \rangle
\end{equation}
Obviously, the sum over all virtual indices is taken in these two expressions, with physical indices remaining open.

The U(1) symmetry in tensor networks is treated so that each of the indices is decomposed into a quantum number (spin projection in this case) and its degeneracy index. For instance, $a_i=\{(m_i,t_i)\}$ in MPS. This decomposition may be understood best if one considers 1D OBC system and follows Ref.~\cite{Schollwoeck2011}. The virtual indices $a_i$ label the states of an appropriate left/right subsystem, which in turn is obtained by adding sites one by one (thus enlarging the system). The number of states covered by a particular $a_i$ is thus determined by the size of the Hilbert space of the subsystem. In case of no symmetries (as in Ref.~\cite{Schollwoeck2011}) the number of states is a power of $(2s+1)$ for any $a_i$. Also it is obvious that physical (spin) indices can be decomposed trivially in the same way: $s_i=\{(s_i,1)\}$.

If the eigenstate of a spin system is U(1) symmetric, the states of the Hilbert space of any subsystem need to be structured with regard to their total spin projection (see also Chapter III of Ref.~\cite{PhysRevB.83.115125}). It is easy for U(1) symmetry, as the basis of the Hilbert space of each spin $i$ is initially chosen as eigenbasis of the spin projection operator. Since the same spin projection of a subsystem can be created by many combinations of spin projections of local spins, the quantum numbers of a virtual (bond) index become degenerate.

In case of total spin projection $S_z=0$ the degeneracies of the quantum numbers at each $a_i$ are multinomial coefficients. Determination of the sets of quantum numbers + their degeneracies for {\it non-trivial} case $S_z \ne 0$ is not obvious, but an appropriate algorithm exists, and the solution for $S_z=0$ is naturally derived as its special case. This is the essence of the next section.

\section{Evaluation of degeneracy sets for OBC}
The MPS for OBC has dummy ends, i.e., leftmost and rightmost virtual indices $a_0$ and $a_N$ have size 1. The way to obtain the global total spin projection $S_z$ is generally outlined in many papers (see, e.g., Refs.~\cite{PhysRevB.83.115125,McCulloch2007}). One needs to `place' the quantum numbers $0$ and $-S_z$ at the ends. I.e., dummy indices $a_0$ and $a_N$ decompose into U(1) symmetric form as: $a_0=\{(0,1)\}$, $a_N=\{(-S_z,1)\}$. (The physical reason for this is that $m_0-m_N=s_1+s_2+\cdots+s_N$ on the one hand and it is required to equal $S_z$ on the other hand)

The task is to determine the degeneracy sets at the virtual indices $a_1, \cdots, a_{N-1}$.
The solution is hidden in the usual back-and-forth `sweeping' DMRG manner~\cite{Schollwoeck2011}. The system is split into left and right parts of $N_L$ and $N_R=N-N_L$ sites, respectively, and, as mentioned before, in the case of U(1) symmetry each part of the system has a set of possible spin projections, $S_z^L$ and $S_z^R$, respectively. They have to satisfy the condition $S_z^L+S_z^R=S_z$, that may exclude some of the $S_z^L$ and $S_z^R$ from the start (for example, if $S_z=N$ for spin-1 system, only $S_z^L=N_L$ and $S_z^R=N_R$ are acceptable).

One can notice that, in fact, $S_z^L=-m_i=-(m_i-m_0)$, and therefore, $S_z^R=S_z-(-m_i)=-(m_N-m_i)$. Obviously, the degeneracy of each $S_z^L$ and $S_z^R$ equals the degeneracy of $m_i$ ($m_0$ and $m_N$ are non-degenerate) and it has be the same. However, it appears that the degeneracies of $m_i$'s created when constructing left part site by site are usually different from those created when constructing right part site by site. Therefore the procedure is: such degeneracies in the left and in the right part must be calculated for each $i$, and their intersection should be taken. This procedure is demonstrated in the Appendix for reader's convenience.

The block structure of the MPS is determined by a Wigner-Eckart theorem, which takes a very simple form in the case of U(1) symmetry. One obtains for MPS that nonzero elements would be for such combinations $s_i,(m_{i-1},t_{i-1}),(m_i,t_i)$ that

\begin{equation}
m_{i-1}=m_i+s_i.
\end{equation}

These conditions plus the determined degeneracy sets for each virtual index are sufficient to construct the MPS (see two examples in Appendix).

\underline{Remark}. Exactly the same algorithm can be applied also to SU(2) symmetric MPS for OBC, except that the representations of the spin projection will be replaced by representations of the total spin.

\section{The algorithm of MPS evaluation}

The previous section represents the most of what U(1) symmetry consideration can provide. The sizes of the MPS tensors at each site as well as the positions of their nonzero elements can be determined. These nonzero elements must be determined explicitly now.

The most advantageous way to do this is to represent the 1D Hamiltonian in the form of matrix product operator (MPO):

\begin{eqnarray}
H=\sum_{\{s,s^{\prime}\}} {\rm Tr}(W^{[1],s_1,s_1^{\prime}} W^{[2],s_2,s_2^{\prime}} \cdots W^{[N],s_N,s_N^{\prime}}) \nonumber \\ |s_1 \cdots s_N \rangle\langle s_1^{\prime} \cdots s_N^{\prime}|
\end{eqnarray}
(explicit example of the MPO for the XXZ model is given in Sec. 6.1 of~\cite{Schollwoeck2011}). Then the MPS elements can be evaluated by optimization algorithm described on page 67 of Ref.~\cite{Schollwoeck2011}. Consecutive minimization of the Lagrangian $L=\langle \psi|H|\psi \rangle - \epsilon \langle \psi|\psi \rangle$ with respect to each conjugated tensor $M^{[i]*}$ leads to the eigenvalue problem:

\begin{equation}
H^{[i]} \nu^{[i]}=\epsilon^{[i]} \nu^{[i]}
\end{equation}
Vector $\nu^{[i]}$ contains the elements of $M^{[i]}$, and the effective Hamiltonian $H^{[i]}$ contains the entire MPO and the MPS elements at all sites except $i$ (which are fixed at the current iteration). One `sweeps' the system back and forth, thus varying MPS parameters site-by-site, and the quantity $\epsilon^{[i]}$ eventually converges to the real eigen-energy.

The aforementioned algorithm is self-contained, therefore I do not review it here. I only refer to two peculiarities that arise for our MPS ansatz:

\begin{enumerate}
\item After the matrix $H^{[i]}$ for the eigenvalue problem is constructed at current site $i$, its rows and columns which correspond to zero elements of $M^{[i]}$ have to be discarded.

\item Left- and right-normalization routine. After new tensor $M^{[i]}$ is obtained from the eigenvalue problem, it is regauged in such way that $Q_L^{[i]}=\sum_{s_i} \, M^{[i],s_i\dag} M^{[i],s_i}=1$ (right sweep) or $Q_R^{[i]}=\sum_{s_i} \, M^{[i],s_i} M^{[i],s_i\dag}=1$ (left sweep), which keeps the wavefunction $|\psi\rangle$ normalized.

Some of the zero entries of the matrix $Q_{L/R}^{[i]}$ are obtained only from zero MPS elements, therefore appropriate equalities are satisfied automatically ($0=0$). Still, the other entries of $Q_{L/R}^{[i]}$ (which are either 1 or 0) impose a number of conditions on nonzero MPS elements they are calculated of. Remarkably, the number of independent MPS elements is thus exactly the same as if the state $|\psi\rangle$ under consideration was represented in the usual form $|\psi\rangle=\sum_{\{s\}} \, c_{s_1,s_2,\cdots,s_N} |s_1 s_2 \cdots s_N\rangle$. This fact clearly shows that the approach is {\it exact}, and it is also demonstrated in Appendix.

\end{enumerate}

\section{The advantages of the approach}

The way how the proposed approach is connected to well-known OBC DMRG implementations (e.g., Ref.~\cite{McCulloch2007}) helps to understand its advantages.

Obviously, the essence of both methods is the same. The thing which differs is the initial MPS ansatz to be optimized. The degeneracy sets are not pre-computed in McCulloch's approach, but they have to be initially set to something. Typically only 2-3 sectors with degeneracy 1 are taken, and such MPS (despite having the required $S_z$) is typically very far from the real eigenstate of the Hamiltonian. The degeneracy sets have to expand throughout the algorithm, otherwise the wrong initial guess makes the algorithm stuck in the local minimum. To avoid this, specific DMRG-like perturbation is applied to the reduced density matrix of the system during MPS regauging. It creates new quantum numbers and changes their degeneracies dynamically throughout the algorithm. Therefore McCulloch's algorithm finally comes to an eigenstate with a good precision (it may be even exact, if the computer can handle the maximal needed MPS size).

Obviously, the crucial quantity that determines the performance of the algorithm is the maximum MPS size $m$. The largest $m$ that can be handled by nowadays computer powers is roughly 20000. Therefore there are two possibilities:

\begin{enumerate}
\item The MPS size that arises in the proposed algorithm can be handled by the computer:

    In this case the proposed algorithm does not need any perturbations (unlike McCulloch's algorithm). The degeneracy sets are known from the start, and they do not need to be changed throughout the algorithm. The MPS is determined exactly, and it happens already within one sweep. McCulloch's algorithm also determines the eigenstate exactly, but it requires bigger number of sweeps and larger computational time. The time ratio for two algorithms was checked for various cases, and it varied from 1.5 to 4.

\item The MPS size that arises in the proposed algorithm exceeds the `critical' number (this happens in a large enough system):

    In this case the MPS size has to be cut in both algorithms. The perturbation to the reduced density matrix becomes crucial in both implementations (our `cut' guess for the degeneracy sets may also be `not very good'). This perturbation enables to judge which quantum numbers to keep (i.e., which are most relevant) and with what degeneracies (the degeneracies of some of the kept quantum numbers have to be reduced). I conclude from my experience as well as the experience of other authors that the `relevant' degeneracy sets become strongly dependent on the Hamiltonian under consideration (roughly said, on the value(s) of the control parameter(s)). The timing of both approaches becomes approximately the same.
\end{enumerate}

\section{Test of the algorithm}

\begin{table*}
\begin{center}
\caption{The energies $E$ of the lowest eigenstates in $S_z=0$ and $S_z=1$ sectors of a spin-1/2 XXZ chain of 12 sites. The relative errors $\frac{|\Delta E|}{E_T}=|\frac{E-E_T}{E_T}|$ with $E_T$ the results of exact diagonalization are also shown. The results indicate the correctness of the proposed algorithm. Note: not energy per site but absolute energy is presented.}
\label{tab1}       
\begin{tabular}{lllllll}
\hline\noalign{\smallskip}
$\Delta$ & $S_z$ & $E$ & $|\Delta E|/E_T$ & $S_z$ & $E$ & $|\Delta E|/E_T$\\
\noalign{\smallskip}\hline\noalign{\smallskip}
-1.5 & 0 & -3.5660045862287 & $2.9 \cdot 10^{-13}$  & 1 & -3.5660584980874 & $3.5 \cdot 10^{-13}$ \\
-1   & 0 & -2.75 & 0 & 1 & -2.75 & 0 \\
0    & 0 & -3.6481149052794 & $3.1 \cdot 10^{-14}$ & 1 & -3.5275782250239 & $3.2 \cdot 10^{-14}$ \\
1    & 0 & -5.1420906328405 & $2.2 \cdot 10^{-14}$ & 1 & -4.8611479370363 & 0 \\
1.5  & 0 & -6.0501401477104 & 0 & 1 & -5.6734344275326 & $2.0 \cdot 10^{-14}$ \\
\noalign{\smallskip}\hline
\end{tabular}
\end{center}
\end{table*}

\begin{table*}
\caption{The energies $E$ of the lowest eigenstates in $S_z=0$ and $S_z=1$ sectors of a spin-1 XXZ chain of 8 sites. The relative errors $|\frac{\Delta E}{E_T}|=|\frac{E-E_T}{E_T}|$ with $E_T$ the results of exact diagonalization are also shown. The results indicate the correctness of the proposed algorithm. Note: not energy per site but absolute energy is presented.}
\label{tab2}       
\begin{center}
\begin{tabular}{lllllll}
\hline\noalign{\smallskip}
$\Delta$ & $S_z$ & $E$ & $|\Delta E|/E_T$ & $S_z$ & $E$ & $|\Delta E|/E_T$\\
\noalign{\smallskip}\hline\noalign{\smallskip}
-1.5 & 0 & -8.2704792249323 & $8.4 \cdot 10^{-12}$ & 1 & -8.2741234280340 & $1.6 \cdot 10^{-12}$ \\
-1   & 0 & -7 & 0 & 1 & -7 & 0 \\
0    & 0 & -8.0204395954077 & $1.4 \cdot 10^{-14}$ & 1 & -7.9059982071590 & $1.4 \cdot 10^{-14}$ \\
1    & 0 & -10.124637222358 & $1.1 \cdot 10^{-14}$ & 1 & -9.9227585483187 & $1.5 \cdot 10^{-13}$ \\
1.5  & 0 & -12.409314862832 & $1.8 \cdot 10^{-14}$ & 1 & -11.689887124543 & 0 \\
\noalign{\smallskip}\hline
\end{tabular}
\end{center}
\end{table*}

The algorithm was checked on a XXZ model with OBC

\begin{equation}
H=\sum_{i=1}^{N-1} \, (s_x^i \otimes s_x^{i+1}+s_y^i \otimes s_y^{i+1}+\Delta\,s_z^i \otimes s_z^{i+1})
\end{equation}
with $s_x,s_y,s_z$ the spin-$s$ SU(2) matrix representations (in case of $s=1/2$ they are Pauli matrices divided by 2). $\Delta$ is called the anisotropy parameter.
%

The tests were performed on small systems, so that the obtained results can be easily checked by exact diagonalization. Namely, I took spin-1/2 XXZ chain of 12 sites (largest possible MPS size is 64) and spin-1 XXZ chain of 8 sites (largest possible MPS size is 81). The appropriate energies of the lowest $S_z=0$ and $S_z=1$ states in the range $-1.5 \le \Delta \le 1.5$ are given in Tables 1, 2. The results clearly indicate the correctness of the approach. As a matter of fact, many of the relative errors presented there are extremely tiny but negative numbers, therefore they are most probably `numerical noise'. It is also checked that the variance per site $\delta H=(\langle\psi|H^2|\psi\rangle-(\langle\psi|H|\psi\rangle)^2)/N$ that shows the `eigen-ness' of the state is always on the level of `numerical noise'.

It should again be noted that the degeneracy sets {\it do not} depend on the Hamiltonian in most cases. The (advantageous) deviation from this rule was found only in the vicinity of a highly entangled point $\Delta=-1$. The reason is: at $\Delta=-1$ many coefficients $c_{s_1,s_2,\cdots,s_N}$ in the expansion $|\psi\rangle=\sum_{\{s\}} \, c_{s_1,s_2,\cdots,s_N} |s_1 s_2 \cdots s_N\rangle$ have integer ratio. Therefore, such a state is represented by effectively less parameters than the general $S_z=0$ state. Accordingly, some elements that were supposed to be nonzero become 0 after optimization. This does not change the sets of quantum numbers, but all their degeneracies effectively reduce to 1 in this case.

\section{Difference between OBC and PBC}

In the case of periodic boundary conditions the calculations are often made using MPS for OBC, just adding one term to the Hamiltonian. On the other hand, special ansatz was introduced~\cite{PhysRevLett.93.227205} for this purpose, with all the virtual indices equal in rights. Now $a_0=a_N$ again, but these indices are not dummy. Therefore it seems somewhat nonphysical to attach different quantum numbers to these indices (unlike OBC). Instead there will be usually more than one quantum number, but the same degeneracy set must be attached to $a_0=a_N$ (let us call it a `boundary index').

The construction and operation of U(1) symmetric MPS for PBC is discussed in detail in Ref.~\cite{PhysRevB.93.054417}. However, the degeneracy sets are chosen by `trial and error' there. If the choice is `bad', the convergence is slow and the precision of the result is badly affected. The algorithm presented here (when properly adapted to PBC) would help to make this choice `wisely'. This would speed-up the convergence significantly (by reduction of the number of optimization steps) and simultaneously improve the precision. It will be demonstrated in a further publication~\cite{Rakov-unpublished}. I only stress the main differences between MPS for OBC and MPS for PBC in the current manuscript:

\begin{enumerate}
\item Boundary quantum number is non-degenerate and has well-defined value (0 or $-S_z$) for OBC, while boundary quantum numbers and their degeneracies must be prescribed for PBC,
\item The quantum numbers in OBC are well-defined for any virtual index, while in PBC they can be calculated only up to a constant. Thus, in MPS for OBC the quantum numbers carry physical meaning, while in MPS for PBC they do not,
\item In MPS for OBC the degeneracy sets grow from the edges to the centre, thus they are unequal. In MPS for PBC unequal degeneracy sets appear only in non-translation invariant Hamiltonians,
\item A specific issue for fermionic systems. In MPS for OBC the quantum numbers of the virtual indices have step 1 (e.g., {-3/2, -1/2, 1/2, 3/2}), but in MPS for PBC numerical experience shows that the step must be 1/2 (e.g., {-1, -1/2, 0, 1/2, 1}).
\end{enumerate}

\section{Difference between 1D (MPS) and 2D (PEPS) cases}

The 2D quantum Hamiltonians with continuous symmetries were investigated, e.g., in Refs.~\cite{PhysRevB.83.125106,PhysRevB.95.235107}. Infinite-size systems were discussed in both publications, and the degeneracy sets were determined by trial and error.
Imaginary-time evolution was applied for calculation of local tensors instead of the variational algorithm in~\cite{PhysRevB.95.235107}.

Explicit treatment of continuous symmetries in 2D quantum Hamiltonians offers substantial speed-up of the calculations, as it reduces the number of parameters within the same PEPS size. In this section I outline basic ideas for the future publication on this issue~\cite{Rakov-unpublished}. Again, the task is to determine the degeneracy sets for the virtual indices. Let us discuss U(1) symmetric PEPS by example of 2D {\it finite-size} square lattice with OBC.

The `selection rule' for the 5-index combination

$s_{i,j},a_{i-1,(j)},a_{i,(j)},a_{(i),j-1},a_{(i),j}$ is

$$
m_{(i),j-1}+m_{i-1,(j)}=s_i+m_{(i),j}+m_{i,(j)}.
$$
with dummy indices $a_{0,(1)}=a_{(1),0}=a_{0,(N)}=a_{(1),N}=a_{(N),0}=a_{N,(1)}=a_{(N),N}=a_{N,(N)}=1$.

It is obvious from Fig. 1(b,c) that $2 \times 2$ lattice with OBC is equivalent to a 1D system of 4 sites with PBC. Its PEPS can therefore be handled similarly to MPS for PBC. The degeneracy sets for virtual dimensions are easy to calculate here, and they slightly depend on the Hamiltonian under consideration. In this simple case one can also use the largest degeneracy set which occurs in the OBC system of size 4.

The situation becomes more complicated for lattice size larger than 2. Total spin projection $S_z$ was ensured by attaching quantum numbers 0 and $-S_z$ to two boundary virtual indices in 1D. These numbers have to be split in two equal portions in 2D. Therefore, one should attach quantum number 0 to the left as well as to the bottom virtual index of the (corner) bottom left site ($i=j=1$) and quantum number $-S_z/2$ to the right as well as to the top virtual index of the (corner) top right site ($i=j=N$). I.e., $m_{(1),0}=m_{0,(1)}=0$, $m_{(N),N}=m_{N,(N)}=-S_z/2$.
The algorithm for the determination of the degeneracy sets combines the features of 1D OBC and 1D PBC cases, and this will be discussed elsewhere~\cite{Rakov-unpublished}.

\begin{figure}
  \includegraphics{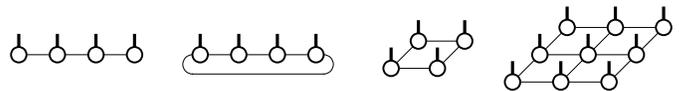}
\caption{Tensor networks discussed in this manuscript. (a) MPS for OBC. (b) MPS for PBC. (c) $2 \times 2$ PEPS for OBC on a square lattice (which is equivalent to MPS for PBC). (d) Larger PEPS for OBC on a square lattice.
The physical (spin) indices are marked in bold. Local MPS tensors obviously have 3 dimensions while PEPS have 5.}
\label{tns}       
\end{figure}


\section{Conclusions}

The algorithm for calculation (from first principles) of the sets of symmetry sectors for virtual indices of
U(1) symmetric matrix product states (MPS) for open boundary conditions is described. Principal difference between open (OBC) and periodic (PBC) boundary conditions is stressed. Similar algorithm for PBC will be outlined in a further publication. The extension to 2D case, i.e., to U(1) symmetric projected entangled pair states (PEPS), appears to be more like in MPS for PBC.

Exactness and correct performance of the algorithm is demonstrated on the eigenstates of spin-1/2 and spin-1 systems with various spin projections $S_z$. It can be checked both analytically and numerically that the predicted degeneracy sets are sufficient to calculate any Hamiltonian of such systems up to 12-14 significant digits. Two explicit examples of the MPS construction are given in the Appendix for reader's convenience.

\section{Appendix}

\subsection{Example of symmetry sector construction: spin-1/2 case}

The MPS for a spin-1/2 system of $N=6$ sites with total spin projection $S_z=1$
is constructed here. According to the algorithm, we set a priori:
$a_0=\{(0,1)\}$, $a_6=\{(-S_z,1)\}=\{(-1,1)\}$.

Let us now construct all possibilities by adding sites one by one from the left, starting from $a_0=$ $\{(0,1)\}$.
Each site can give $\pm 1/2$, so we get:

\noindent
$a_1=\{(-1/2,1),(1/2,1)\}$, $a_2=\{(-1,1),(0,2),(1,1)\}$,

\noindent
$a_3=\{(-3/2,1),(-1/2,3),(1/2,3),(3/2,1)\}$,

\noindent
$a_4=\{(-2,1),(-1,4),(0,6),(1,4),(2,1)\}$,

\noindent
$a_5=\{(-5/2,1),(-3/2,5),(-1/2,10),(1/2,10),(3/2,5),$

\noindent
$(5/2,1)\}$, $a_6=\{(-3,1),(-2,6)$, $(-1,15),(0,20),(1,15),$

\noindent
$(2,6),(3,1)\}$.

Let us now construct all the possibilities by adding sites one by one from the right, starting from $a_6=\{(-1,1)\}$.
We get:

\noindent
$a_5=\{(-3/2,1),(-1/2,1)\}$, $a_4=\{(-2,1),(-1,2),(0,1)\}$,

\noindent
$a_3=\{(-5/2,1),(-3/2,3),(-1/2,3),(1/2,1)\}$,

\noindent
$a_2=\{(-3,1),(-2,4),(-1,6),(0,4),(1,1)\}$,

\noindent
$a_1=\{(-7/2,1),(-5/2,5),(-3/2,10),(-1/2,10),(1/2,5),$

\noindent
$(3/2,1)\}$, $a_0=\{(-4,1),(-3,6),(-2,15),(-1,20),(0,15),$

\noindent
$(1,6),(2,1)\}$.

The intersection gives the needed degeneracy sets: $a_0=\{(0,1)\}$
(in line with a priori setting),

\noindent
$a_1=\{(-1/2,1),(1/2,1)\}$, $a_2=\{(-1,1),(0,2),(1,1)\}$,

\noindent
$a_3=\{(-3/2,1),(-1/2,3),(1/2,1)\}$,
$a_4=\{(-2,1),(-1,2),(0,1)\}$, 
$a_5=$ $\{(-3/2,1),(-1/2,1)\}$,

\noindent
$a_6=\{(-1,1)\}$ (again - in line with a priori setting).

In accordance to the selection rule $m_{i-1}=s_i+m_i$
one obtains the block structure of MPS matrices $M^{[i],s_i}$ as follows:

{\small
$$
M^{[1],+1/2}=\left(\begin{array}{cc} x & 0 \end{array}\right), M^{[1],-1/2}=\left(\begin{array}{cc} 0 & x \end{array}\right);
$$

$$
M^{[2],+1/2}=\left(\begin{array}{cccc} x & 0 & 0 & 0 \\ 0 & x & x & 0 \end{array}\right), M^{[2],-1/2}=\left(\begin{array}{cccc} 0 & x & x & 0 \\ 0 & 0 & 0 & x \end{array}\right);
$$

$$
M^{[3],+1/2}=\left(\begin{array}{ccccc} x & 0 & 0 & 0 & 0 \\ 0 & x & x & x & 0 \\ 0 & x & x & x & 0 \\ 0 & 0 & 0 & 0 & x \end{array}\right), M^{[3],-1/2}=\left(\begin{array}{ccccc} 0 & x & x & x & 0 \\ 0 & 0 & 0 & 0 & x \\ 0 & 0 & 0 & 0 & x \\ 0 & 0 & 0 & 0 & 0 \end{array}\right);
$$

$$
M^{[4],+1/2}=\left(\begin{array}{cccc} x & 0 & 0 & 0 \\ 0 & x & x & 0 \\ 0 & x & x & 0 \\ 0 & x & x & 0 \\ 0 & 0 & 0 & x \end{array}\right), M^{[4],-1/2}=\left(\begin{array}{cccc} 0 & x & x & 0 \\ 0 & 0 & 0 & x \\ 0 & 0 & 0 & x \\ 0 & 0 & 0 & x \\ 0 & 0 & 0 & 0 \end{array}\right);
$$

$$
M^{[5],+1/2}=\left(\begin{array}{cc} 0 & 0 \\ x & 0 \\ x & 0 \\ 0 & x \end{array}\right), M^{[5],-1/2}=\left(\begin{array}{cc} x & 0 \\ 0 & x \\ 0 & x \\ 0 & 0 \end{array}\right);
$$

$$
M^{[6],+1/2}=\left(\begin{array}{c} 0 \\ x \end{array}\right), M^{[6],-1/2}=\left(\begin{array}{c} x \\ 0 \end{array}\right).
$$
}
Here, `{\it x}' denote nonzero matrix elements (which are usually different even inside the same matrix). These nonzero elements are determined in the optimization procedure outlined in Section 4.

It is interesting to show that the number of independent MPS parameters exactly covers the necessary basis to construct the state with required $S_z=1$. We have 4 spins `up' and 2 spins `down', therefore there are 15 relevant basis states $|s_1 s_2 \cdots s_N\rangle$. The normalization condition $\langle\psi|\psi\rangle=1$ makes the number of independent parameters $15-1=14$.

Our MPS ansatz contains 42 nonzero elements. After the optimization all 6 MPS tensors $M^{[i]}$ are right-normalized. One can calculate the number of conditions that would make $Q_R^{[i]}=\sum_{s_i} \, M^{[i],s_i} M^{[i],s_i\dag}=1$ satisfied for $i=1$ through 6. For our MPS ansatz the matrices $Q_R^{[i]}$ are

{\small
$$
Q_R^{[1]}=\left(\begin{array}{c} x \end{array}\right); Q_R^{[2]}=\left(\begin{array}{cc} x & 0 \\ 0 & x \end{array}\right); Q_R^{[3]}=\left(\begin{array}{cccc} x & 0 & 0 & 0 \\ 0 & x & x & 0 \\ 0 & x & x & 0 \\ 0 & 0 & 0 & x \end{array}\right);
$$

$$
Q_R^{[4]}=\left(\begin{array}{ccccc} x & 0 & 0 & 0 & 0 \\ 0 & x & x & x & 0 \\ 0 & x & x & x & 0 \\ 0 & x & x & x & 0 \\ 0 & 0 & 0 & 0 & x \end{array}\right); Q_R^{[5]}=\left(\begin{array}{cccc} x & 0 & 0 & 0 \\ 0 & x & x & 0 \\ 0 & x & x & 0 \\ 0 & 0 & 0 & x \end{array}\right); Q_R^{[6]}=\left(\begin{array}{cc} x & 0 \\ 0 & x \end{array}\right).
$$
}

There are $1+2+6+11+6+2=28$ conditions overall. The number of independent MPS parameters is therefore $42-28=14$ as expected.

\subsection{Example of symmetry sector construction: spin-1 case}

The MPS for a spin-1 system of $N=4$ sites with total spin projection $S_z=2$
is constructed here. According to the algorithm, we set a priori: $a_0=\{(0,1)\}$, $a_4=\{(-S_z,1)\}=\{(-2,1)\}$.

Let us calculate all possible combinations by adding sites one by one from the left,
starting from $a_0=\{(0,1)\}$.
Each site can give $-1,0,1$, so we get: $a_1=\{(-1,1),$

\noindent
$(0,1),(1,1)\}$, $a_2=\{(-2,1),(-1,2),(0,3),(1,2),(2,1)\}$,

\noindent
$a_3=\{(-3,1),(-2,3),(-1,6),(0,7),(1,6),(2,3),(3,1)\}$,

\noindent
$a_4=\{(-4,1),(-3,4)$, $(-2,10),(-1,16),(0,19),(1,16),$

\noindent
$(2,10),(3,4),(4,1)\}$.

Let us calculate all possible combinations by adding sites one by one from the right,
starting from $a_4=\{(-2,1)\}$.
We get: $a_3=\{(-3,1),(-2,1),(-1,1)\}$,

\noindent
$a_2=\{(-4,1),(-3,2),(-2,3),(-1,2),(0,1)\}$,

\noindent
$a_1=\{(-5,1),(-4,3),(-3,6),(-2,7),(-1,6),(0,3),$

\noindent
$(1,1)\}$,
$a_0=\{(-6,1),(-5,4),(-4,10),(-3,16),(-2,19),$

\noindent
$(-1,16),(0,10),(1,4),(2,1)\}$.

The intersection gives appropriate sets for each virtual index:
$a_0=\{(0,1)\}$ (in line with a priori setting), $a_1=\{(-1,1),(0,1),(1,1)\}$,
$a_2=\{(-2,1),(-1,2),$ $(0,1)\}$, $a_3=\{(-3,1),(-2,1),(-1,1)\}$,
$a_4=\{(-2,1)\}$ (again - in line with a priori setting).

In accordance to the selection rule $m_{i-1}=s_i+m_i$
one obtains the block structure of MPS matrices $M^{[i],s_i}$ as follows:

{\small
$$
M^{[1],+1}=\left(\begin{array}{ccc} x & 0 & 0 \end{array}\right), M^{[1],0}=\left(\begin{array}{ccc} 0 & x & 0 \end{array}\right), M^{[1],-1}=\left(\begin{array}{ccc} 0 & 0 & x \end{array}\right);
$$

$$
M^{[2],+1}= \left(\begin{array}{cccc} x & 0 & 0 & 0 \\ 0 & x & x & 0 \\ 0 & 0 & 0 & x \end{array}\right), \hspace{0.5cm} M^{[2],0}= \left(\begin{array}{cccc} 0 & x & x & 0 \\ 0 & 0 & 0 & x \\ 0 & 0 & 0 & 0 \end{array}\right),
$$

$$
M^{[2],-1}= \left(\begin{array}{cccc} 0 & 0 & 0 & x \\ 0 & 0 & 0 & 0 \\ 0 & 0 & 0 & 0 \end{array}\right); \hspace{0.5cm} M^{[3],+1}= \left(\begin{array}{ccc} x & 0 & 0 \\ 0 & x & 0 \\ 0 & x & 0 \\ 0 & 0 & x \end{array}\right),
$$

$$
\hspace{0.2cm} M^{[3],0}= \left(\begin{array}{ccc} 0 & x & 0 \\ 0 & 0 & x \\ 0 & 0 & x \\ 0 & 0 & 0 \end{array}\right), \hspace{0.5cm} M^{[3],-1}= \left(\begin{array}{ccc} 0 & 0 & x \\ 0 & 0 & 0 \\ 0 & 0 & 0 \\ 0 & 0 & 0 \end{array}\right);
$$

$$
M^{[4],+1}= \left(\begin{array}{c} 0 \\ 0 \\ x \end{array}\right),
M^{[4],0}= \left(\begin{array}{c} 0 \\ x \\ 0 \end{array}\right),
M^{[4],-1}= \left(\begin{array}{c} x \\ 0 \\ 0 \end{array}\right).
$$
}

Again `{\it x}' denote nonzero matrix elements to be determined in the optimization procedure outlined in Section 4.

Let us show again that the number of independent MPS parameters covers the necessary basis to construct the state with required $S_z=2$. In this case we have either 3 spins with $s_i=1$ and 1 spin with $s_i=-1$, or 2 spins with $s_i=1$ and 2 spins with $s_i=0$. Therefore there are 10 possible basis states $|s_1 s_2 \cdots s_N\rangle$, and the  normalization condition $\langle\psi|\psi\rangle=1$ reduces the number of independent parameters by 1: $10-1=9$.

There are 22 nonzero elements in our MPS ansatz. After the optimization all 4 MPS tensors $M^{[i]}$ are right-normalized. The conditions $Q_R^{[i]}=\sum_{s_i} \, M^{[i],s_i} M^{[i],s_i\dag}=1$ must be satisfied for $i=1$ through 4. For our MPS ansatz the matrices $Q_R^{[i]}$ are

{\small
$$
Q_R^{[1]}=\left(\begin{array}{c} x \end{array}\right); Q_R^{[2]}=\left(\begin{array}{ccc} x & 0 & 0 \\ 0 & x & 0 \\ 0 & 0 & x \end{array}\right);
$$

$$
Q_R^{[3]}=\left(\begin{array}{cccc} x & 0 & 0 & 0 \\ 0 & x & x & 0 \\ 0 & x & x & 0 \\ 0 & 0 & 0 & x \end{array}\right); Q_R^{[4]}=\left(\begin{array}{ccc} x & 0 & 0 \\ 0 & x & 0 \\ 0 & 0 & x \end{array}\right).
$$
}

It gives $1+3+6+3=13$ equations overall. The number of independent MPS parameters is therefore $22-13=9$ as expected.

\begin{acknowledgements}
Useful discussions with Ian P. McCulloch are highly appreciated. The author also thanks Physikalisch-Technische Bundesanstalt (Braunschweig, Germany) for provision of {\it Mathematica} software during guest visits.
\end{acknowledgements}



\end{document}